\documentclass[a4paper,twocolumn]{article}
\usepackage{IWAC2026_template_latex}
\usepackage{amsmath}
\usepackage{amssymb}
\usepackage{amsfonts}
\usepackage{overcite}
\usepackage{arydshln}
\usepackage{bm}
\usepackage{newtxtext,newtxmath}
\usepackage[affil-it]{authblk}
\usepackage[setpagesize=false]{hyperref}
\usepackage{booktabs}
\usepackage{fancyhdr}
\usepackage{enumitem}
\usepackage{caption}
\usepackage[top=24.2truemm,bottom=21.8truemm,left=14.1truemm,right=19.9truemm]{geometry}
\usepackage{dblfloatfix}
\usepackage{graphicx}
\usepackage{xcolor}
\usepackage{verbatim}
\usepackage{float}

\setlist[itemize]{topsep=-1ex, parsep=-1ex}
\setlist[enumerate]{topsep=-1ex, parsep=-1ex}
\makeatletter
\def\fnum@figure{Fig. \thefigure. }
\def\fnum@table{Table \thetable. }
\makeatother

\makeatletter 
\def\@citess#1{\textsuperscript{#1)}} 
\makeatother
\usepackage{setspace}

\makeatletter
\renewcommand{\@biblabel}[1]{#1)}   
\makeatother

\makeatletter
\def\section{\@startsection{section}{1}{\z@}{0.8ex plus 1.0ex minus 0ex}{0.8ex plus 1ex minus 0ex}{\normalsize\bf}}
\def\subsection{\@startsection{subsection}{2}{\z@}{0.0ex plus -0.1ex minus 0.0ex}{0.1ex plus -0.5ex}{\normalsize\bf}}
\def\subsubsection{\@startsection{subsubsection}{3}{\z@}{0.5ex plus 0.5ex minus 0.2ex}{0.5ex plus 0.2ex}{\normalsize\bf}}
\makeatother

\makeatletter
\def\@listi{%
  \setlength{\leftmargin}{\leftmargini}%
  \setlength{\parsep}{0pt}%
  \setlength{\topsep}{0.5\baselineskip}%
  \setlength{\itemsep}{0pt}%
}
\makeatother

\renewcommand{\headrulewidth}{0pt}

\title{\fontsize{14pt}{14pt}\selectfont Aircraft and Fleet Sizing for Regional Air Mobility: College Town Case Studies}

\author[1)]{\normalsize Jung Ho Park}
\author[1)]{\normalsize Changyeob Lee}
\author[1)]{\normalsize Shangqing Cao}
\author[1)]{\normalsize Raja Sengupta}
\author[1)]{\normalsize Mark Hansen}
\author[2)]{\normalsize Pavan Yedavalli}
\affil[1)]{\footnotesize Department of Civil and Environmental Engineering, University of California, Berkeley, USA}
\affil[2)]{\footnotesize Wisk Aero, USA}
\abstract{
We examine how aircraft seat configuration interacts with daily operation in Regional Air Mobility by applying a joint supply-demand optimization framework that simultaneously determines market share, fare, and flight schedule. The framework integrates a binary logit discrete choice model into a task assignment formulation, capturing passengers' mode choice between Regional Air Mobility and driving across spatiotemporal origin-destination pairs. We evaluate three U.S. college town corridors under 4-, 6-, and 8-seat configurations across cost scales from 0.4 to 1.0 and fleet sizes from 12 to 30 aircraft. Profitability and throughput serve as primary performance metrics, and we analyze pricing power, operating cost, and revenue to explain performance variation across markets. We find that larger aircraft configurations and fleet sizes do not improve profitability universally. Larger aircraft are preferred where economies of scale are favorable and demand is sufficient and directionally balanced. The best configuration in these case studies is the 4-seat in imbalanced markets and the 6-seat in balanced or dense markets.
}


\begin{document}

\maketitle
\thispagestyle{fancy}
\normalsize

\setlength{\abovedisplayskip}{-2ex}
\setlength{\belowdisplayskip}{-0.5ex}
\setlength{\abovedisplayshortskip}{-2ex}
\setlength{\belowdisplayshortskip}{-0.5ex}

\section{Introduction}

The aviation industry is experiencing a paradigm shift with the advent of hybrid Vertical Take-Off and Landing (VTOL) technologies. Unlike traditional fixed-wing aircraft that require massive infrastructure, VTOLs operate with a unique, flexible footprint. This opens up the Regional Air Mobility (RAM) market (typically 100--500 kilometers), targeting demand that has historically been underserved by commercial airlines \cite{nasa2021ram}. The recent shift in focus to hybrid VTOL further enables vehicle configurations with higher payload capacities and faster turnaround times.

Despite this technological advantage, operators and planners still struggle to identify viable early-adopter markets, which is especially critical for aircraft Original Equipment Manufacturers (OEMs) who commit to seat capacity and other configuration choices at the preliminary design stage \cite{roy2018next}. The industry currently lacks empirical case studies that quantify the viability of these niche markets. Analyzing the RAM market differs from analyzing intra-city travel in that trips are longer and a substantial portion is spent on high-speed freeways. This makes market health sensitive to local geography and daily congestion, both of which vary widely across candidate corridors.

Major U.S. college towns are of particular interest as early-adopter markets for RAM. They host highly mobile populations of students, faculty, alumni, and visiting families, often disconnected from major aviation hubs by 1- to 3-hour drives to the nearest commercial airport. We focus on three corridors that span the operational diversity these markets exhibit. The contributions of this paper are threefold. 1) A framework for evaluating how aircraft seat configuration, fixed costs, and operating costs affect downstream operations, intended to inform aircraft OEMs and operators at the preliminary design stage. 2) Application of a joint supply-demand optimization framework to three U.S. college town corridors as RAM case studies. 3) Quantification and sensitivity analysis of RAM profitability and throughput across seat configurations, fleet sizes, and operating cost projections.

\begin{figure*}[t!]
    \centering
    \includegraphics[width=0.95\textwidth]{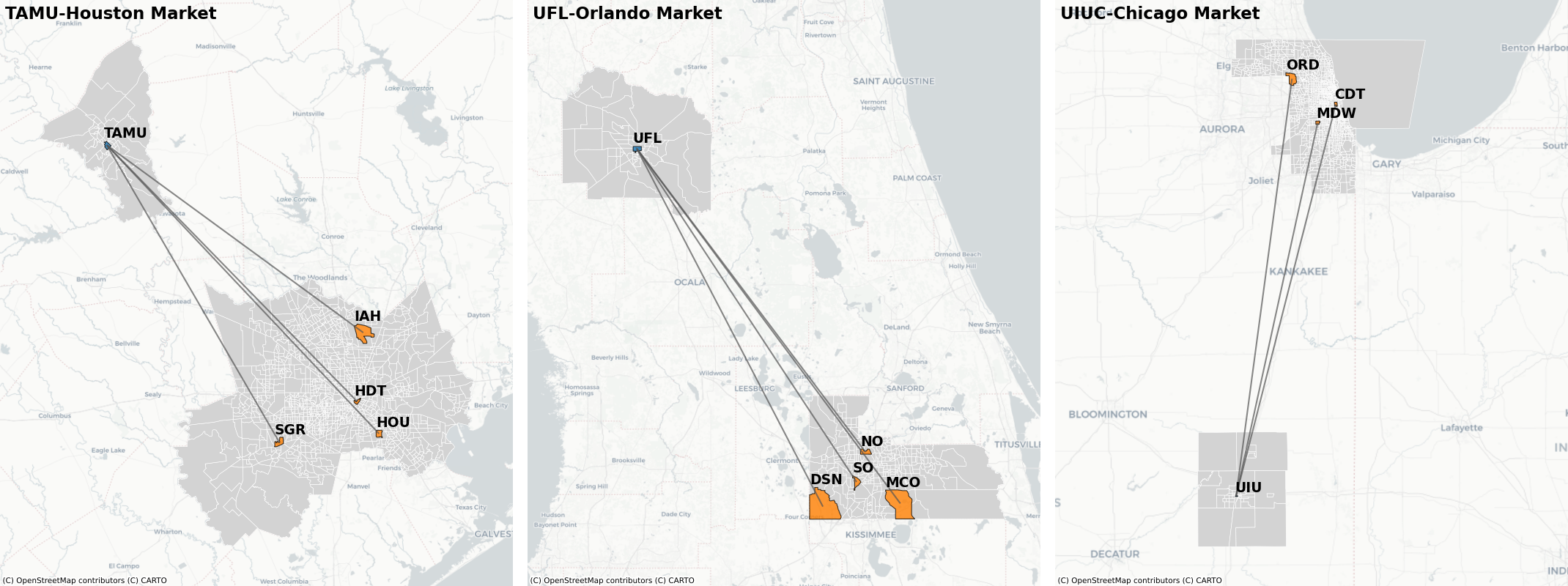}
    \caption{Vertiport networks for the three case-study corridors.}
    \label{fig:vertiport_network}
\end{figure*}

\section{Methodology} 
\label{sec:methodology}

\subsection{Service concept}
\label{sec:service_concept}

We model RAM as a middle-mile service in which a fleet of hybrid VTOL aircraft is dispatched from a network of vertiports to serve passenger demand between Origin-Destination (OD) pairs. Flight schedules are generated dynamically by the joint supply-demand optimization framework. For tractability, we assume that passenger demand is served at the intended time of departure, that gates and parking pads are not capacity-constrained, and that the fleet operates homogeneously with a single aircraft configuration.

We index each flight leg by $i \in \mathcal{I}$, where $i = (i_o, i_d, t_{i_o}, t_{i_d})$ encodes the origin vertiport, destination vertiport, departure time, and arrival time. The corresponding flight time is $t_i^{flight} = t_{i_d} - t_{i_o}$. From the passenger's perspective, a RAM trip on leg $i$ consists of a first-mile access leg by ride-hailing, in-terminal processing, the flight itself, and a last-mile egress leg by ride-hailing, giving a total door-to-door time of $t^{RAM}_i = t^{acc}_i + t^{proc} + t^{flight}_i + t^{egr}_i$. The corresponding total trip cost $r^{RAM}_i$ comprises the ride-hailing fares for access and egress and the RAM fare charged by the operator, which is itself a decision variable.

The competing alternative on each OD pair is private driving, with travel time $t^{drive}_i$ and total cost $r^{drive}_i$ reflecting per-mile operating expense and any destination-specific parking fees. Passengers choose between RAM and driving according to a binary logit discrete choice model whose utility is linear in total time and total cost, with the cost coefficient scaled by the local Value of Time (VoT) to capture spatial heterogeneity in willingness to pay. We focus on the binary competition between RAM and driving because private automobiles and air travel together account for over 97\% of U.S. long-distance travel for trips exceeding 50 miles \cite{rasmidatta2006mode}, and driving remains the dominant mode within the under-400-mile range that covers most RAM operations \cite{moeckel2015mode}.

\subsection{Task assignment problem}
\label{sec:network_flow}
% \vspace{2.5ex}

We model the daily RAM operation as a network flow problem on a time-expanded vertiport network. Time is discretized into windows $t \in \{1, \dots, T\}$, and the set of flight legs $\mathcal{I}$ populates the spatiotemporal arcs. The decision variable $x_i \in \mathbb{Z}_{\geq 0}$ denotes the number of aircraft dispatched on leg $i$. Repositioning between legs is captured by $y_{(i,j)} \in \mathbb{Z}_{\geq 0}$ on the set of valid arcs $\mathcal{R} = \{(i,j) \mid t_{i_d} + f_{(i_d, j_o)} \leq t_{j_o}\}$, where $f_{(i_d, j_o)} = 0$ when $i_d = j_o$. Letting $s_i$ denote the number of aircraft sourced from the depot to begin leg $i$, fleet feasibility is enforced through flow conservation, $\sum_{i} s_i \leq K$ for total fleet size $K$, and

\begin{equation}
\sum_{j \in \mathcal{I}} y_{(j,i)} + s_i 
= x_i + \sum_{j \in \mathcal{I}} y_{(i,j)} 
\quad \forall\; i \in \mathcal{I}.
\end{equation}

We impose a Final Approach and Take-Off (FATO) capacity constraint at each vertiport $v \in \mathcal{V}$ during each time window $t$. Defining $\mathcal{I}^{dep}_{v,t}$ and $\mathcal{I}^{arr}_{v,t}$ as the sets of scheduled departures and arrivals, and $\mathcal{R}^{dep}_{v,t}$ and $\mathcal{R}^{arr}_{v,t}$ as the corresponding repositioning sets, for all $v$ and $t$,

\begin{equation}
\sum_{i \in \mathcal{I}^{dep}_{v,t} \cup \mathcal{I}^{arr}_{v,t}} x_i 
+ \sum_{(i,j) \in \mathcal{R}^{dep}_{v,t} \cup \mathcal{R}^{arr}_{v,t}} y_{(i,j)} 
\leq \text{Cap}_{v,t}
\end{equation}

The operating cost of each flight has a fixed flight-cycle component $c_{FC}$ covering takeoff, landing, and vertiport fees, and a variable flight-hour component $c_{FH}$ scaling with flight duration. The total operating cost across revenue and repositioning flights is

\begin{equation}
\begin{aligned}
C = & \sum_{i \in \mathcal{I}} x_i 
\left( c_{FC} + c_{FH} \cdot f_{(i_o, i_d)} \cdot \tfrac{\Delta t}{60} \right) \\
& + \sum_{(i,j) \in \mathcal{R}} y_{(i,j)} 
\left( c_{FC} + c_{FH} \cdot f_{(i_d, j_o)} \cdot \tfrac{\Delta t}{60} \right)
\end{aligned}
\end{equation}

\noindent where $\Delta t$ is the time window duration in minutes.

\subsection{Mode choice integration and joint optimization}
\label{sec:joint_optimization}
\vspace{1.5ex}

% Existing work on RAM and UAM operations has largely treated demand generation and operational optimization as separate steps. Supply-side studies focus on fleet sizing, scheduling, and repositioning under fixed or exogenous demand \cite{cao_fleet_2024, roy_flight_2022, kotwicz_herniczek_fleet_2024, husemann_analysis_2024}. Demand-side studies use mode choice models to estimate UAM market share and willingness to pay \cite{fu_exploring_2019, ilahi_understanding_2021, boddupalli_mode_2024, coppola_urban_2024, brunelli_sp_2023, hae_choi_exploring_2022}, with travel time savings and Value of Time consistently identified as the dominant utility drivers. The closest RAM-specific work, by Justin et al.\cite{justin2021demand, justin2022regional}, jointly optimizes fleet composition and flight schedules over the U.S. Northeast Corridor for aircraft ranging from 9 to 48 seats, but demand in their formulation is a pre-computed input with static fares. We instead adopt the joint supply-demand optimization framework of Cao et al.\cite{cao_jointsupplydemand_2025}, originally developed for UAM airport access at JFK, which determines fare and market share jointly with operational decisions.

Existing work on RAM and UAM operations has largely treated demand generation and operational optimization as separate steps. Supply-side studies focus on fleet sizing, scheduling, and repositioning under fixed or exogenous demand \cite{cao_fleet_2024, roy_flight_2022, kotwicz_herniczek_fleet_2024, husemann_analysis_2024}. Demand-side studies use mode choice models to estimate UAM market share and willingness to pay \cite{fu_exploring_2019, ilahi_understanding_2021, boddupalli_mode_2024, coppola_urban_2024, brunelli_sp_2023, hae_choi_exploring_2022}, with travel time savings and Value of Time consistently identified as the dominant utility drivers. The closest RAM-specific work, by Justin et al.\cite{justin2021demand, justin2022regional}, jointly optimizes fleet composition and flight schedules over the U.S. Northeast Corridor for different seat configurations, but demand in their formulation is a pre-computed input with static fares. Recent work has begun to address the supply-demand feedback through joint optimization \cite{lv_urban_2024, jin_integrated_2024, kirste_modeling_2024}, but typically with fares fixed or handled heuristically. The fundamental challenge of joint optimization is the nonconvexity, which can be mitigated by market-share inversion technique \cite{dong_dynamic_2009, li_pricing_2011}. We adopt the joint supply-demand optimization framework of Cao et al.\cite{cao_jointsupplydemand_2025}, which integrates market-share inversion into a discrete fleet-routing framework for UAM airport access at JFK, and apply it to the RAM context.

We now integrate the moode choice model directly into the optimization, so that the market share of RAM (and hence the fare) on each leg $i$ is determined jointly with the dispatching decisions $x_i$ and $y_{(i,j)}$. The systematic utility of each alternative $k \in \{RAM, drive\}$ is $V^k_i = \vartheta^t_i \cdot t^k_i + \vartheta^r_i \cdot r^k_i$. To incorporate spatial heterogeneity in willingness to pay, the cost coefficient scales with the local Value of Time as $\vartheta^r_i = \beta_r \cdot \text{VoT}_i / 60$, where $\beta_r$ is the reference cost coefficient and $\text{VoT}_i$ is expressed in \$/hr. The RAM market share $\omega^{RAM}_i$ follows the standard binary logit form over $V^{RAM}_i$ and $V^{drive}_i$. The objective maximizes daily operating profit, total revenue minus operating cost.

\begin{equation}
\max_{\omega^{RAM}_i,\; r^{RAM}_i,\; x_i,\; y_{(i,j)}} \quad 
\sum_{i \in \mathcal{I}} r^{RAM}_i \cdot d_i \cdot \omega^{RAM}_i \;-\; C
\end{equation}

\noindent where $d_i$ is the total addressable market on leg $i$. Optimizing directly over the fare $r^{RAM}_i$ produces a non-concave objective because of the logit probability function. We therefore apply the market-share inversion technique, which re-expresses the fare as a function of the market share decision variable using the IIA property of the logit model.

\begin{equation}
r^{RAM}_i = \frac{1}{\vartheta^r_i} \left[ 
\ln \omega^{RAM}_i - \ln(1 - \omega^{RAM}_i) 
+ V^{drive}_i - \vartheta^t_i \cdot t^{RAM}_i 
\right]
\end{equation}

The resulting objective contains entropy terms $\omega^{RAM}_i \ln \omega^{RAM}_i$ and $\omega^{RAM}_i \ln(1 - \omega^{RAM}_i)$ and a bilinear ratio $\omega^{RAM}_i / x_i$, which are linearized via piecewise-linear approximation and McCormick envelope constraints, yielding a Mixed-Integer Linear Program (MILP). We refer the reader to Cao et al.\ \cite{cao_jointsupplydemand_2025} and Kim et al.\ \cite{kim_strategic_2025} for the complete derivation and linearization scheme.

\section{Case studies: US college towns}
\label{sec:case_studies}
% Describe why college towns

College towns offer a particularly relevant setting for evaluating early-stage RAM viability. They host highly mobile populations of students, faculty, alumni, and visiting families who travel frequently between campus and major metropolitan areas. These campuses are typically located 1.5 to 2.5 hours by car from the nearest major aviation hub, with limited commercial air service and few alternative modes beyond driving. A hybrid VTOL service connecting a campus vertiport directly to metropolitan nodes can offer meaningful time savings at a scale appropriate to this demand, but the economic viability of such a service depends on the specific characteristics of each corridor.

\subsection{Market selection and characterization}

We selected three U.S. college town corridors for our case studies: 1) Texas A\&M University (TAMU) – Houston, 2) University of Florida (UFL) – Orlando, and 3) University of Illinois Urbana-Champaign (UIUC) – Chicago. These pairs were chosen to represent diverse regional characteristics where campus locations are typically positioned 1.5 to 2.5 hours away from major metropolitan destinations by car.

TAMU -- Houston (large, short-haul) is the largest of the three markets by total daily demand at approximately 9,571 trips/day, with a driving time of roughly 95 minutes between the campus and the Houston metropolitan area. Demand is moderately imbalanced, with college-to-city flows about 35\% larger than the reverse.

UIUC -- Chicago (mid-sized, balanced, long-haul) is a directionally balanced market of approximately 4,406 trips/day, separated by the longest driving distance of the three corridors at around 150 minutes.

UFL -- Orlando (small, imbalanced) is the smallest market at approximately 3,904 trips/day and the most directionally imbalanced, with college-to-city demand roughly twice the reverse flow. The driving distance is approximately 116 minutes.

Table~\ref{table:case_studies} summarizes the average weekday directional demand and reference driving times for each corridor. We revisit these archetypes throughout the results section, where they correspond to qualitatively different operational regimes.

\begin{table}[t!]
    \footnotesize
    \caption{Key characteristics showing average weekday's directional demand volumes from Replica, average travel distances and in-vehicle travel times (IVTT) in each mode for the selected college town corridors.}
    \centering
    \resizebox{1.05\columnwidth}{!}{
    \begin{tabular}{lcccccc}
        \toprule
        City Pair & Demand & \multicolumn{2}{c}{Driving} & \multicolumn{2}{c}{RAM} \\
        \cmidrule(lr){3-4} \cmidrule(lr){5-6}
        (Origin -- Dest.) & (trips/day) & Dist. (mi) & Time (min) & Dist. (mi) & Time (min) \\
        \midrule
        TAMU -- Houston & 4,980 & 95.5 & 102 & 84.5 & 57 \\
        Houston -- TAMU & 3,540 & 95.7 & 100 & 84.5 & 60 \\
        \addlinespace
        UFL -- Orlando  & 1,130 & 120.6 & 115 & 100.0 & 61 \\
        Orlando -- UFL  & 2,430 & 118.4 & 114 & 100.0 & 61 \\
        \addlinespace
        UIUC -- Chicago & 2,444 & 135.4 & 137 & 126.7 & 68 \\
        Chicago -- UIUC & 1,962 & 136.9 & 141 & 126.7 & 70 \\
        \bottomrule
    \end{tabular}
    }
    \label{table:case_studies}
\end{table}

\subsection{Data and parameters}
\label{sec:parameters}
% Replica, demand numbers, VoT, cost table, cost scaling factor

\subsubsection{Travel demand data}
We characterize passenger demand using Replica\cite{Replica2026}, a high-resolution activity-based data platform that synthesizes Location-Based Services (LBS) traces, consumer demographics, and land-use data into a representative synthetic population. Each modeled trip includes origin, destination, departure and arrival times, trip purpose, primary mode, and traveler socio-economic attributes. Using raw connected-car or LBS data directly would expose the analysis to low penetration rates and sampling biases, while the synthetic population preserves disaggregated trip-level detail without these confounds. For each corridor we extract average weekday demand at the catchment-area level for both directions. The Replica dataset also provides door-to-door driving times that serve as the baseline for the mode choice model, and corridor-specific ride-hailing travel times for the first-mile and last-mile legs of a RAM trip.

\begin{figure}[h]
    \centering
    \includegraphics[width=0.8\columnwidth]{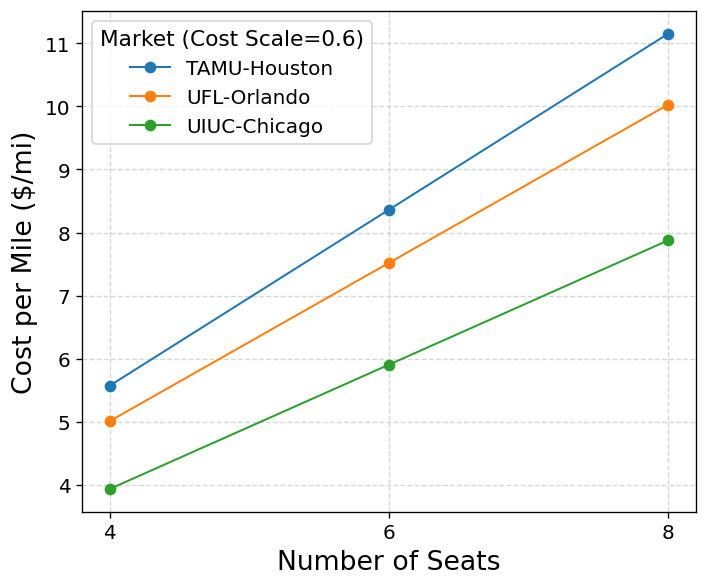}
    \caption{Average operating cost per mile by aircraft configuration in each of the three markets at cost scale 0.6.}
    \label{fig:RAM_cost_per_mile}
\end{figure}

\subsubsection{Service parameters}
Each RAM trip includes a fixed in-terminal processing time of $t^{proc} = 10$ minutes covering security screening and boarding. Ride-hailing is assumed to incur a total wait time of 10 minutes across the first and last mile, split evenly. The airborne flight time is computed using publicly available flight mechanics equations for hybrid VTOL configurations \cite{emin_evtol_2024, yedavalliAssessingValueUrban2021}. For the mode choice model, we adopt a Value of Time of \$36 per hour, drawn from estimates for non-business travelers in prior UAM studies \cite{rimjha_urban_2021}. This rate reflects the price-sensitivity of typical college-town travelers, who are predominantly students and staff rather than premium business segments.

\subsubsection{Driving and RAM cost}
The driving cost is modeled as a per-mile operating cost of \$0.64 \cite{bts2021household}, with an additional \$20 destination parking fee \cite{parkopedia2022global} applied only to city-heading trips.

Operating costs for the hybrid VTOL are synthesized from a range of industry sources, including manufacturer projections (Wisk), aviation training school operating metrics, and institutional reports from IATA and the American Academy of Actuaries. We separate costs into per Flight Hour (FH) and per Flight Cycle (FC) components. As a reference point, we use \$113/FH per seat and \$64/FC per seat, respectively. These numbers are based on projection for 4-seat hybrid VTOL. OPEX/FH include pilot labor, maintenance, insurance, and fuel. OPEX/FC include component replacement and landing fees. The cost landscape for hybrid VTOL operations carries substantial uncertainty at this stage.

% \begin{table}[!hbtp]
%     \footnotesize
%     \caption{Projected per-flight operating cost (hybrid VTOL) in USD, expressed per Flight Hour (FH) and per Flight Cycle (FC).}
%     \label{table:operational_cost}
%     \centering
%     \begin{tabular}{lccc}
%         \toprule
%         Operating Metric & 4-Seat Min & 4-Seat Mean & 4-Seat Max \\
%         \midrule
%         Pilot Labor (\$/FH) & 120 & 150 & 180 \\
%         Maintenance (\$/FH) & 140 & 250 & 450 \\
%         Insurance (\$/FH) & 90 & 160 & 280 \\
%         Fuel and Energy (\$/FC) & 90 & 160 & 280 \\
%         Battery Replacement (\$/FC) & 70 & 150 & 300 \\
%         Landing Fees (\$/FC) & 45 & 70 & 120 \\
%         Per-Seat OPEX/FH & 136 & 150 & 191 \\
%         Per-Seat OPEX/FC & fix & fix & fix \\
%         \bottomrule
%     \end{tabular}
% \end{table}

\subsubsection{Cost scaling}
To represent economies of scale across seat configurations of 4, 6, and 8 seats, we apply a multiplicative cost scaling factor to the mean per-seat OPEX of the 4-seat baseline. We assume that the same scaling factor is applied to both per flight-hour and per flight-cycle components. While the two cost types do not scale together in reality, this simplification reduces the combinatorial space of simulations and provides a tractable representation of economoies of scale.  As an example, if increasing seat capacity reduces operating costs by 10\% per step, an analyst can compare cost scales of 0.7, 0.6, and 0.5 for 4-, 6-, and 8-seat aircraft respectively. This formulation allows the model to capture how larger aircraft can potentially lower the per-seat-mile cost, directly influencing the service's pricing power and the subsequent modal shift predicted. It also enables the representation of different operational regimes at varying economies of scale. 

Figure~\ref{fig:RAM_cost_per_mile} shows the resulting cost per mile for each aircraft configuration across the three markets at a cost scale of 0.6. The cost per mile is highest in markets with shorter flight distances, where the fixed per-cycle cost is spread over fewer miles.  It is worth noting that hybrid propulsion systems are still at an early stage of development; the absolute cost values include large uncertainties, and it is therefore more beneficial to analyze the system using the scale factor rather than using the absolute cost values. In the results that follow, we sweep cost scales from 0.4 to 1.0 to capture this full range of operational assumptions, and treat the question of which seat configuration is most profitable as a robustness question across cost levels rather than a point comparison.

\subsection{Vertiport network}
\label{sec:network_design}

For each corridor we configure a vertiport network from the high-volume trip clusters identified in the Replica data. Each college town is served by a single campus vertiport that aggregates the geographically concentrated demand originating in or destined for the campus area. The metropolitan side is served by multiple vertiports placed at high-demand nodes including major airports and central business districts. This asymmetric configuration reflects the qualitative difference between the campus catchment (geographically concentrated) and the metropolitan catchment (geographically dispersed), and it reduces the first-mile and last-mile burden on the metropolitan side. Figure~\ref{fig:vertiport_network} shows the resulting networks. The metropolitan nodes are Houston (HDT, SGR, IAH, HOU), Orlando (NO, SO, DSN, MCO), and Chicago (CDT, ORD, MDW). The IAH, MCO, and ORD nodes correspond to the primary commercial airports in each region and are included in part to capture the substantial airport-access component of college-town demand.

\section{Results and analysis}
\label{sec:results}

We analyze the results from four different perspectives to advise on aircraft designs through the lens of service profitability, market penetration, and pricing power. Section~\ref{sec:profitability} examines daily operating profit across markets, seat configurations, and cost scales as the direct outcome. Section~\ref{sec:throughput} measures passenger throughput as a complementary indicator of market penetration. Section~\ref{sec:pricing_power} decomposes the profit into pricing power and unit economics and explains why markets with similar demand can perform very differently. Section~\ref{sec:oem_implications} summarizes implications for aircraft OEMs at the early design stage.

\subsection{Profitability}
\label{sec:profitability}

\begin{figure*}[t!]
    \centering
    \includegraphics[width=\textwidth]{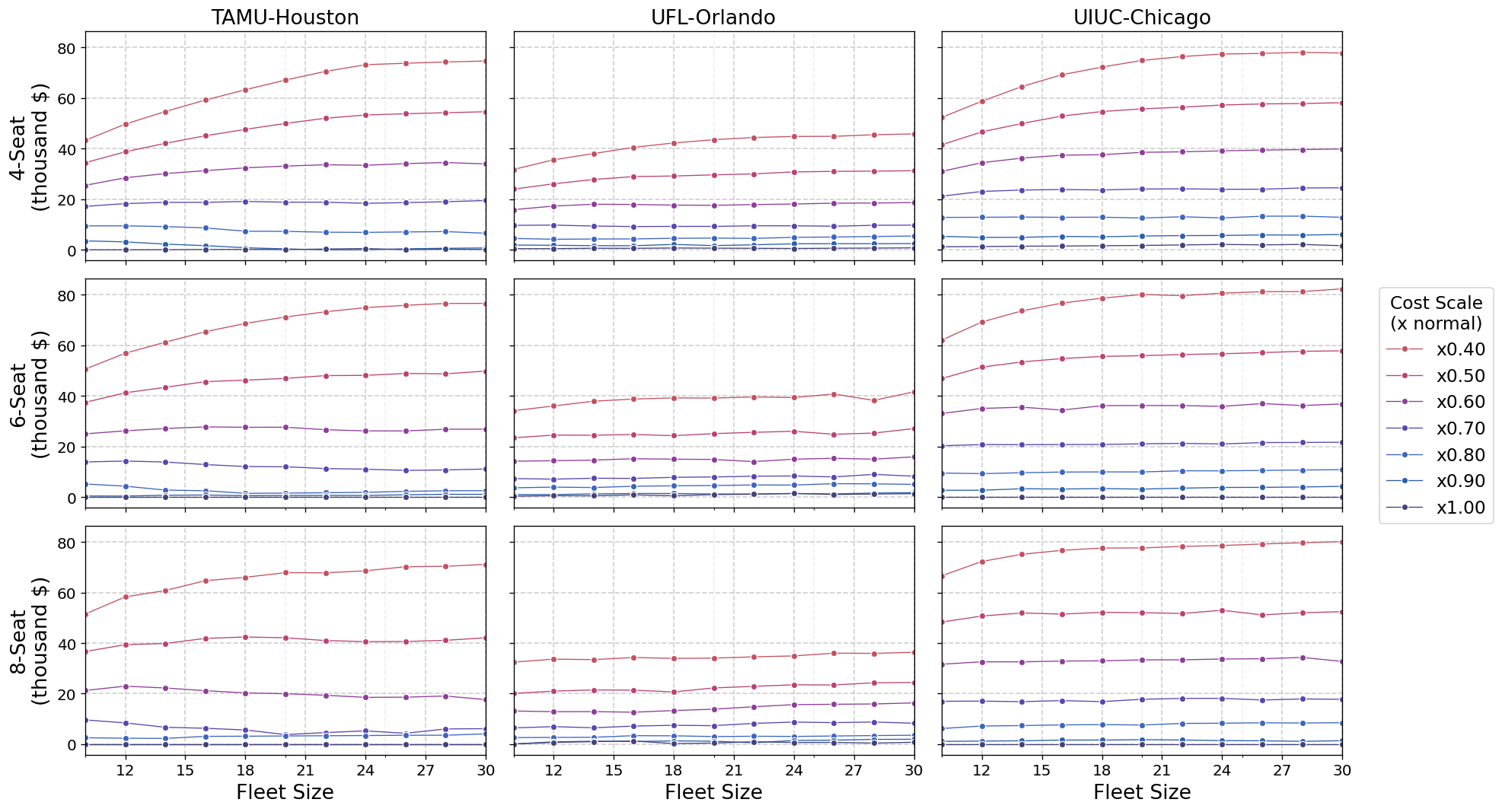}
    \caption{Daily operating profit across the three case-study markets as a function of fleet size, seat configuration, and cost scale.}
    \label{fig:operating_cost_summary}
\end{figure*}

Figure~\ref{fig:operating_cost_summary} shows daily operating profit across the three case-study markets as a function of fleet size, seat configuration, and cost scale. 

Profitability has a nonlinear relationship with the cost scaling factor. Cost reductions yield much larger profit gains in low-cost regimes than in high-cost regimes. In the TAMU-Houston market, dropping the cost scale from 60\% to 40\% roughly doubles operating profit in absolute terms, while a similar step from 100\% to 80\% barely changes the profitability. The same pattern holds in the other two markets. This shows that profitability is much more sensitive to cost reduction at different stages. The cost need to be scaled down significantly to start this service. But it will generally lead to much higher profitability when economy of scale is achieved. 

Larger fleet size does not consistently improve profitability, especially in high-cost regimes. Profit curves either plateau or stay nearly flat across the fleet-size range at high cost scales. This is because higher operating costs disincentivize frequent service and repositioning. The system must operate fewer flights at higher fares than fill more flights at lower margins. 

The three markets exhibit qualitatively different profit profiles at the same cost scale, even when their total demand is comparable. UIUC-Chicago consistently outperforms the other two, and UFL-Orlando underperforms substantially despite having roughly the same total daily demand as UIUC-Chicago. TAMU-Houston, the largest market by demand, generates similar operating profit to UIUC across most fleet and cost configurations. These differences point to market characteristics beyond raw demand volume. We examine each market in turn.

TAMU-Houston (large, short-haul) is the largest market by total demand (9,571 trips/day) and the most cost-pressured by flight distance as shown in Fig.~\ref{fig:RAM_cost_per_mile}. As a result, operators have less headroom to lower fares to capture demand. The FATO capacity constraint compounds this in the medium-to-high cost regime, where profit plateaus at smaller fleet sizes because adding aircraft would push vertiport operations beyond the per-window cap. Despite the cost handicap, sheer demand volume sustains TAMU-Houston at profit levels comparable to UIUC-Chicago. The 6-seat aircraft outperforms the other configurations at lower cost scales, where the steady dense demand fills the additional seats.

UFL-Orlando (small, imbalanced) is the smallest market (3,904 trips/day) and the most directionally imbalanced, with college-to-city flows approximately double the reverse. Repositioning empty aircraft to balance the directional flows is much more costly in RAM because every empty leg incurs the same fixed flight-cycle cost as a revenue leg, over substantially longer distances.Operating profit is substantially lower than the other two markets across every fleet size, cost scale, and seat configuration. The 4-seat aircraft performs best here because its flexibility matches the modest demand on the lighter direction and limits the cost penalty of imbalanced flows.

UIUC-Chicago (mid-sized, balanced, long-haul) is the strongest-performing market across the board. Total demand is comparable to UFL-Orlando, but flows are directionally balanced and flight distances are the longest of the three corridors at approximately 130 miles. The long flights amortize fixed per-cycle costs over more miles, producing the lowest per-mile operating cost of the three markets (See Fig.~\ref{fig:RAM_cost_per_mile}). Operators on this corridor can offer competitive fares while generating good profit margins. 6- and 8-seat aircraft outperforms at low cost scales, which suggests that markets that does not require operational flexibilities prefer higher seat capacity aircraft.

\begin{figure*}[t!]
    \centering
    \includegraphics[width=\textwidth]{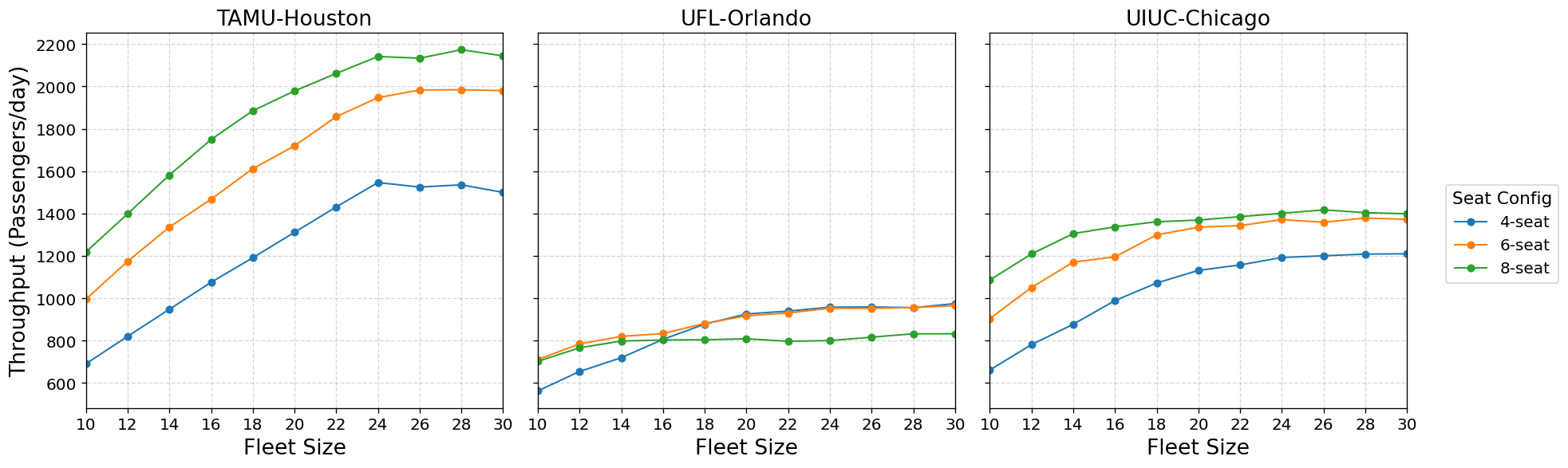}
    \caption{Daily passenger throughput at cost scale 0.6, by market, seat configuration, and fleet size.}
    \label{fig:throughput_c60}
\end{figure*}

\subsection{Throughput and market penetration}
\label{sec:throughput}
\vspace{1.5ex}

The total passengers served provide important insights as it represents the market penetration rate of the RAM service. Figure~\ref{fig:throughput_c60} shows daily passenger throughput at a 60\% cost scale across the three markets, by seat configuration and fleet size.

The plot interestingly shows the diminishing return on throughput with increasing fleet size. TAMU-Houston plateaus at the highest throughput, around 2,000 passengers per day for the 8-seat configuration, with the fleet size around 24 aircraft. UIUC-Chicago plateaus earlier, around 18 aircraft, at a slightly lower level of about 1,400 passengers per day across the larger seat configurations. UFL-Orlando saturates earliest and at the lowest level. Each saturation point identifies the operationally appropriate fleet size for that market, which can take into consideration both throughput and operating profit.

\begin{figure}[!htbp]
    \centering
    \includegraphics[width=\columnwidth]{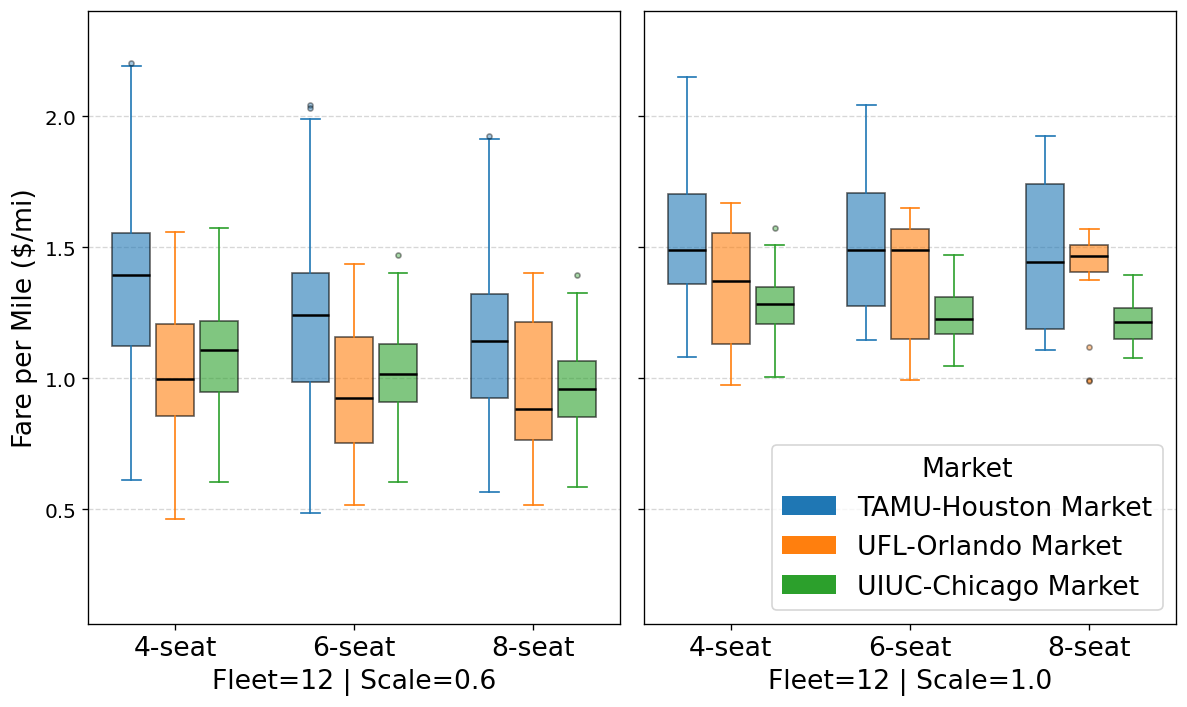}
    \caption{Per-mile fare distribution by market and seat configuration at cost scales 0.6 (left) and 1.0 (right), fleet size 12.}
    \label{fig:fare_per_mile}
\end{figure}

There is also interaction between seat configuration and market. In TAMU-Houston, throughput climbs monotonically with seat capacity, as expected. Each flight serves more passengers, and the dense market can operate with more supply of aircraft. In UIUC-Chicago, the throughput gap different seat aircraft is narrower, because long flight times limit the number of revenue cycles per aircraft per day and the additional seats may only partially filled. In UFL-Orlando the 8-seat configuration produces lower throughput than the 6-seat. This is another evidence of the configuration penalty in imbalanced markets. Larger aircraft are forced into less frequent service to maintain load factors, and the lower frequency in turn raises the generalized cost to passengers, reducing the RAM market share and total throughput. Hence, bigger aircraft can serve fewer total passengers in a thin imbalanced market. For an OEM, this means that the appropriate measure of market penetration depends on examining aircraft configuration with market characteristics. A 4-seat aircraft will penetrate the UFL-style markets more effectively than a 6- or 8-seat will, even though the larger aircraft serve more passengers per flight.

\subsection{Pricing power and unit economics}
\label{sec:pricing_power}
\vspace{1ex}

We refer to the operator's ability to charge fares above the marginal cost of serving a passenger as pricing power. 

\begin{figure*}[t!]
    \centering
    \includegraphics[width=\textwidth]{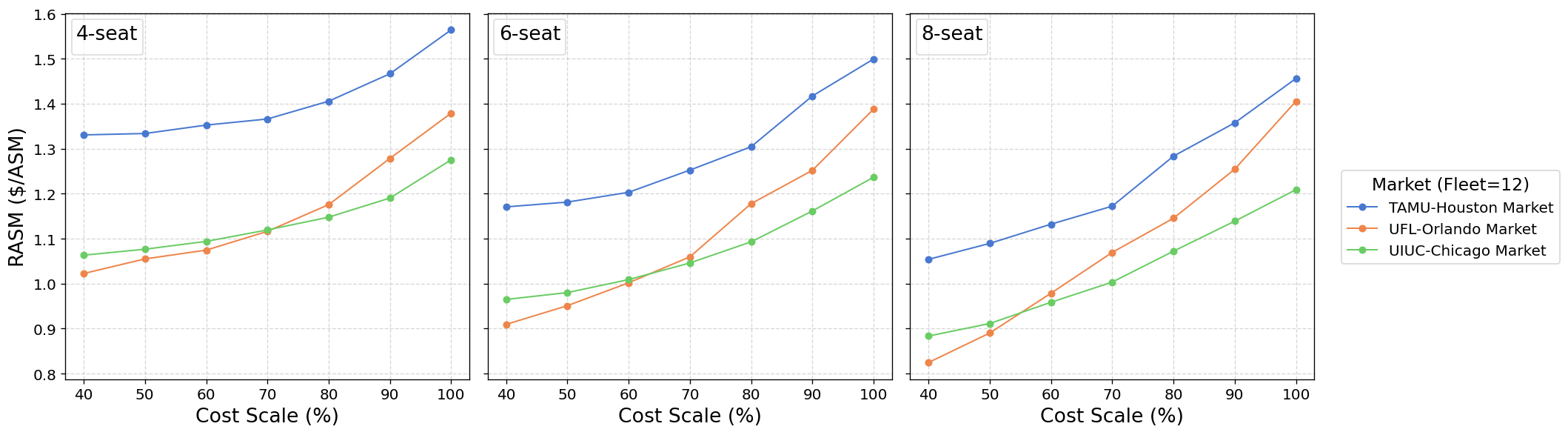}
    \caption{Revenue per Available Seat Mile (RASM) as a function of cost scale, by market and seat configuration, at fleet size 12.}
    \label{fig:rasm_cost}
\end{figure*}

In our setup, pricing power on a given corridor depends on 1) the time advantage of RAM over driving, since passengers will pay more when the time saving is large, 2) demand density, since dense markets are less elastic to fare increases, 3) the local willingness to pay, which we capture through Value of Time. Figure~\ref{fig:fare_per_mile} shows the per-mile fare distributions across the three markets and three seat configurations at two cost scales, and Figure~\ref{fig:rasm_cost} shows the corresponding unit revenue, Revenue per Available Seat Mile (RASM), as a function of cost scale at a fleet size of 12.

At a low cost scale (0.6), operators have meaningful fare flexibility. TAMU-Houston shows the widest fare distribution, indicating that operators can exercise dynamic pricing strategies across different times and trips. Higher-capacity aircraft allow operators to offer lower per-mile fares to capture more demand. UFL-Orlando shows the narrowest distribution and the lowest median fare across all configurations, reflecting the limited pricing power of an imbalanced market with small demand. UIUC-Chicago is in between the two other markets. At a high cost scale (1.0), the fare distributions compress upward into a narrow band across all markets and configurations. The cost level forces operator into a similar pricing scheme for all seat configurations regardless of underlying market strength.

Figure~\ref{fig:rasm_cost} clarifies this explanation. TAMU-Houston produces the highest RASM across every cost scale and seat configuration, confirming the most pricing power per unit of seat-mile sold. UIUC-Chicago has consistently lower RASM, and UFL-Orlando is lowest at low cost scales but rises steeply as cost scale increases, crossing UIUC across all seat configurations. As costs rise, UFL operators cannot raise fares enough to maintain market share against driving, so most of the demand collapses to a small batch of passengers who are willing to pay premium for the small time saving. Comparing RASM with the per-mile operating cost in Fig.~\ref{fig:RAM_cost_per_mile}, TAMU's high RASM is paired with the highest cost per mile, due to shorter flight. The margin is therefore not as wide as the RASM alone suggests. UIUC has moderate RASM but the lowest cost per mile, giving it the widest margin per seat-mile and the highest total profit. UFL suffers from lower RASM and high cost per mile combination. 

Pricing power therefore shows both advantages and disadvantages of a market. Non-flexible markets with high cost of operation give small pricing power to operators. Operators do not gain more flexibilities in setting the fare by utilizing higher seat aircraft. When the cost of operation scales down, we see more benefits of utilizing higher seat aircraft, especially for a balanced market with sufficient demand. 

\subsection{Implications for vehicle design}
\label{sec:oem_implications}
\vspace{1ex}

The three market archetypes encountered in our case studies and the seat configurations that perform best in each under our cost projection are summarized here. For a larger market with high total addressable demand, the 6-seat configuration is preferred at low cost scale, suggesting that economies of scale for larger aircraft may make them strong candidates for regional-scale mobility. The results also show that at the current demand level, starting with a small fleet size is beneficial at an early deployment stage.

The results offer a way to make a decision on the correct seat configuration given some market characteristics. Total demand volume is not a sufficient signal for sizing decisions. While UIUC and UFL have similar daily demand, UFL significantly underperforms UIUC, due to the combination of flight distance, directional balance, and demand density. Markets like TAMU favor the 6-seat configuration in low-cost regimes because steady demand can fill the additional seats with sufficient pricing flexibility. In a very low cost scenario, imbalanced markets like UFL favor the 4-seat configuration due to its operational flexibility, which can match supply to imbalanced demand at finer granularity.

The operating cost of hybrid VTOL is a major source of uncertainty at this early stage. The dominant seat configuration shifts with cost scale in some markets, which means a design decision based on a single cost point may target the wrong market. All three markets we analyzed have moderate demand volume, which makes the 4-seat configuration the most flexible and robust design choice across all scenarios. The picture would shift for denser markets, where capacity limits at vertiports including parking pads, gates, and FATO start to bind. In those settings, RAM service may favor larger aircraft. More broadly, RAM cannot treat vertiport capacity and directional balance as secondary constraints the way short-haul UAM analyses can.

The first-order problem for OEMs and operators is to narrow down the projected operating cost of aircraft. Operating cost can be treated as a range of values, and the decision on seat configuration can be considered as a robustness problem. That is, the goal is to choose the seat configuration that performs best across a range of cost scales and market conditions.

\subsection{Limitations}
\label{sec:limitations}
% \vspace{5ex}

The vertiport capacity model is simplified. While it considers FATO capacity, our formulation assumes that gates and parking pads are unlimited. The joint optimization model is also limited to a binary mode choice between RAM and driving. Expanding the framework to compete with other modes such as public transit would require modification of both the choice model and the demand inputs. The addressed market includes airport access and regular regional trips by car, but does not take into account inter-airport trips by commercial airlines, which represent one of the most significant demand sources for regional aviation. Integrating an airline model into RAM is significantly challenging, as airlines operate under their own revenue maximization schemes and the interaction would require further assumptions about partnership or competition. This is a topic for more principled examination in future work.

\section{Conclusion}
\label{sec:conclusion}

In this study, we examined how aircraft seat configuration interacts with the daily operation of a Regional Air Mobility service across three U.S. college town markets. Using a joint supply-demand optimization framework that simultaneously determines fare and flight schedule, we evaluated 4-, 6-, and 8-seat configurations against fleet sizes from 12 to 30 and cost scales from 0.4 to 1.0. The analysis showed that larger aircraft and fleets do not universally improve profitability. The dominant seat configuration is market-specific and shifts with cost scale. The 4-seat configuration emerges as the most robust default across the moderate-demand markets we analyze, the 6-seat is preferred in dense markets at low cost scales, and structurally imbalanced markets underperform regardless of configuration. Throughput analysis showed that larger aircraft in general increases throughput and market penetration, but it is inefficient in imbalanced markets where the resulting drop in flight frequency raises the generalized cost faced by passengers. This offer insights to aircraft OEMs that flight distance, directional balance, and demand density are the key market characteristics that drive the seat-configuration decisions. Future work should expand the framework to capture inter-airport demand and multi-modal competition with public transit and commercial airlines.

\section*{Acknowledgments}
The authors would like to thank Wisk Aero, LLC, a wholly-owned subsidiary of Boeing, for providing insights, expertise, and reference data.

\bibliographystyle{iwac}
\bibliography{reference}

\end{document}